\documentclass[conference]{IEEEtran}
\IEEEoverridecommandlockouts
\usepackage{cite}
\usepackage{amsmath,amssymb,amsfonts}
\usepackage[ruled,vlined]{algorithm2e}
\usepackage{graphicx}
\usepackage{textcomp}
\usepackage{xcolor}
\usepackage{ulem}
\usepackage{subfigure}
\usepackage{multirow}
\newcommand{\eg}{\textit{e}.\textit{g}., }
\newcommand{\ie}{\textit{i}.\textit{e}.,}
\newcommand{\etal}{\textit{et} \textit{al}.}

\emergencystretch=\maxdimen
\def\BibTeX{{\rm B\kern-.05em{\sc i\kern-.025em b}\kern-.08em
    T\kern-.1667em\lower.7ex\hbox{E}\kern-.125emX}}
\begin{document}

\title{Hi-Gen: Generative Retrieval For Large-Scale Personalized E-commerce Search}




\author{\IEEEauthorblockN{Yanjing Wu\IEEEauthorrefmark{1},
Yinfu Feng\IEEEauthorrefmark{1},
Jian Wang\IEEEauthorrefmark{1},
Wenji Zhou\IEEEauthorrefmark{1},
Yunan Ye\IEEEauthorrefmark{1},
Rong Xiao\IEEEauthorrefmark{1},
Jun Xiao\IEEEauthorrefmark{2}}
\IEEEauthorblockA{\IEEEauthorrefmark{1}Alibaba Group\\
\{qinge.wyj, yinfu.fyf, eric.wj, eric.zwj, yunan.yyn\}@alibaba-inc.com,\, 
xiaorong.xr@taobao.com}
\IEEEauthorblockA{\IEEEauthorrefmark{2}Zhejiang University\\
junx@zju.edu.cn}
}


\maketitle

\begin{abstract}
\sloppy{}
Leveraging generative retrieval (GR) techniques to enhance search systems is an emerging methodology that has shown promising results in recent years. In GR, a text-to-text model maps string queries directly to relevant document identifiers (docIDs), dramatically simplifying the retrieval process. However, when applying most GR models in large-scale E-commerce for personalized item search, we must face two key problems in encoding and decoding. (1) Existing docID generation methods ignore the encoding of efficiency information, which is critical in E-commerce. (2) The positional information is important in decoding docIDs, while prior studies have not adequately discriminated the significance of positional information or well exploited the inherent interrelation among these positions. To overcome these problems, we introduce an efficient \textbf{Hi}erarchical encoding-decoding \textbf{Gen}erative retrieval method (Hi-Gen) for large-scale personalized E-commerce search systems. Specifically, we first design a representation learning model using metric learning to learn discriminative feature representations of items to capture semantic relevance and efficiency information. Then, we propose a category-guided hierarchical clustering scheme that makes full use of the semantic and efficiency information of items to facilitate docID generation. Finally, we design a position-aware loss to discriminate the importance of positions and mine the inherent interrelation between different tokens at the same position. This loss boosts the performance of the language model used in the decoding stage. Besides, we propose two variants of Hi-Gen (\ie Hi-Gen-I2I and Hi-Gen-Cluster) to support online real-time large-scale recall in the online serving process. Extensive experiments on both public and industry datasets demonstrate the effectiveness and efficiency of Hi-Gen. It gets 3.30$\%$ and 4.62$\%$ improvements over SOTA for Recall@1 on the public and industry datasets, respectively. Moreover, deploying Hi-Gen to a large-scale E-commerce platform leads to 6.89$\%$ and 1.42$\%$ improvements over the baseline model in the online AB-testing experiment in terms of RecallNum and Gross Merchandise Value (GMV), respectively. In addition, Hi-Gen beats the basic Differentiable Search Index (DSI) model and BM25 in zero-shot learning scenarios, which proves its generalization capabilities.
\end{abstract}


\section{Introduction}
With the rapid development of online E-commerce in recent years, personalized information retrieval technology has become increasingly important. For example, a user who requests a query on E-commerce websites like Amazon.com and AliExpress.com can quickly retrieve lots of relevant items from different seller stores. Information retrieval not only helps customers find their favorite items but also generates revenue for these business platforms.

Roughly speaking, existing retrieval techniques can be classified broadly into three categories: sparse retrieval (SR)~\cite{lee2022generative,dai2019context}, dense retrieval(DR)~\cite{karpukhin2020dense,yu2021dual, zeng2023,wang2018billion,zhang2020towards}, and generative retrieval (GR)~\cite{tay2022transformer}. SR acquires documents by evaluating the relevance between query terms and document terms. DR encodes queries and documents into high-dimensional vectors with a neural network and computes their similarity to retrieve relevant documents. In GR, the standardized pipelined \textit{retrieve-then-rank} strategy for information retrieval is replaced by a simple \textit{sequence-to-sequence} (Seq2Seq) strategy. The GR models in which generative Language Models (LMs)~\cite{radford2019language,lewis2019bart,liu2019roberta,yao2023deepspeed} play a key role allow for end-to-end training and can serve as a differentiable sub-component in complex neural models for building intelligent E-commerce systems to enhance the customer experience. In light of this, GR has gained widespread attention in recent years.

The whole process of GR consists of a series of decisions, associated with the sub-problems of document representation, indexing, and retrieval. The performance of GR heavily relies on the choices and optimizations made for these sub-problems.

Firstly, in the document representation stage, some term-based and vector-based strategies have been proposed to represent a document~\cite{tay2022transformer, semanticenhance}. For term-based approaches, a few representative words or tokens are selected to describe the semantic content of a document\cite{semanticenhance}. In contrast, vector-based approaches present documents in dense vectors and model their similarities with an inner product or cosine similarity, which effectively solves the problem of similar semantic meanings but different word representations faced by term-based methods\cite{tay2022transformer}. In addition to the semantic information, other side information, such as item conversion efficiency information, is also significant in e-commerce scenarios. Efficiency information represents the efficiency-related indicators of the item, such as the Click-Through Rate (CTR) and Gross Merchandise Value (GMV). This information is very important for improving the efficiency of the personalized e-commerce search system. Nevertheless, there are relatively few works that explicitly learn both the semantic and efficiency information in document\footnote{In the field of E-commerce search, we treat a product as a special document, thus we sometimes use the word \textit{document} or \textit{item} to express a \textit{product} when there is no ambiguity in this paper.} representation. We call this problem \textit{the document representation problem}.


Secondly, document representations are mapped into their corresponding docIDs in the indexing stage. Existing docID generation methods can be roughly classified into three groups: 1) unstructured atomic identifiers, 2) naively structured string identifiers, and 3) semantically structured identifiers. The most naive way to represent documents is to assign each document an arbitrary (and possibly random) unique integer identifier, which is called \textit{unstructured atomic identifiers}~\cite{tay2022transformer}. To enhance the representation ability, a small group of arbitrary unique integers, words, or tokens are used to form docID. We refer to these as \textit{naively structured string identifiers}. To capture the semantic information about documents, a hierarchical clustering process over document embeddings has been employed in~\cite{tay2022transformer} to generate a decimal tree, which guides the generation of \textit{semantically structured identifiers}. Structural representation has a wide range of potential applications because structural relationships exist in many real-world applications. However, the docID generation methods in most GR models ignore encoding crucial efficiency information about items in large-scale E-commerce search scenarios. We refer to this problem as \textit{the docID generation problem}.

Next, the retrieval process in GR is accomplished by decoding docIDs given a query. When docIDs are formed as a series of organized tokens, decoding them is a structured prediction task where the position of each token is very important. A good prediction method should capture dependencies between output components and make joint predictions that respect these dependencies, leading to more accurate predictions compared to making independent predictions for each component. However, prior studies have not adequately discriminated the significance of positional information or well exploited the inherent interrelation among these positions. Recently, Lian \etal~\cite{lian2023personaltm} began to consider the importance of different positions in docID by applying an additional hierarchical loss at the decoding steps, but they ignored the inherent interrelation among different tokens at the same position. Here we use the term \textit{docID decoding problem} to describe this issue.


Finally, directly using GR in online large-scale recall applications would encounter the service delay problem caused by poor inference calculation performance. We call this issue \textit{the large-scale recall problem}. Most previous research on GR only presented findings from small datasets~\cite{fan2023}. There is a lack of real-world application testing results on large-scale applications such as industry E-commerce search systems.



To overcome the above-mentioned problems, we propose an efficient \textbf{Hi}erarchical encoding-decoding \textbf{Gen}erative retrieval method (\textbf{Hi-Gen}). For \textit{document representation problem}, \textbf{Hi-Gen} designs a representation learning model using metric learning to learn discriminative feature representations of items to capture both semantic relevance and efficiency information. To overcome \textit{docID generation problem}, our method proposes a two-layer category-guided hierarchical clustering scheme, which facilitates the docID-related encoding and decoding process. As for \textit{docID decoding problem}, a position-aware loss is proposed to boost the performance of the language model used in the decoding stage. Furthermore, we propose two variants of Hi-Gen (\ie Hi-Gen-I2I and Hi-Gen-Cluster) to solve \textit{the large-scale recall problem} in real large-scale industry applications. Overall, the main contributions of our work are:

\begin{itemize} 
        \item We introduce a novel method for docID generation in GR. Our method first designs a representation learning model along with metric learning to learn a discriminative feature representation of a document for maintaining both semantic relevance and efficiency information. Subsequently, the category-guided hierarchical clustering scheme is adopted to make full use of the semantic and efficiency information of documents in docID generation.

        \item The position-aware loss is proposed to enhance the performance of the language model in the decoding stage. It aims to discriminate the importance of different positions and mine the semantic and efficiency differences among various tokens at the same position.
        
        \item We propose two variants of Hi-Gen (\ie \textit{Hi-Gen-I2I} and \textit{Hi-Gen-Cluster}) to support online real-time large-scale recall in the online service process. Hi-Gen-I2I rapidly expands generative retrieval results with item-to-item (I2I) recall algorithms, which are commonly used in modern search and recommendation systems. In contrast, the Hi-Gen-Cluster method exploits the hierarchical structure information encoded in docID by truncating the decoded result early in decoding to bring in large-scale recall.
        
        \item In comparison to the original DSI, our methods achieve state-of-the-art performance across both public and industrial large-scale datasets. Furthermore, the practical usefulness of our methods in real-world search systems is evidenced by the improvements seen in online AB experiment metrics on a large-scale E-commerce platform.
\end{itemize}

\section{Related Work}
\sloppy{}
\begin{figure*}
    \centering
    \includegraphics[width=\linewidth]{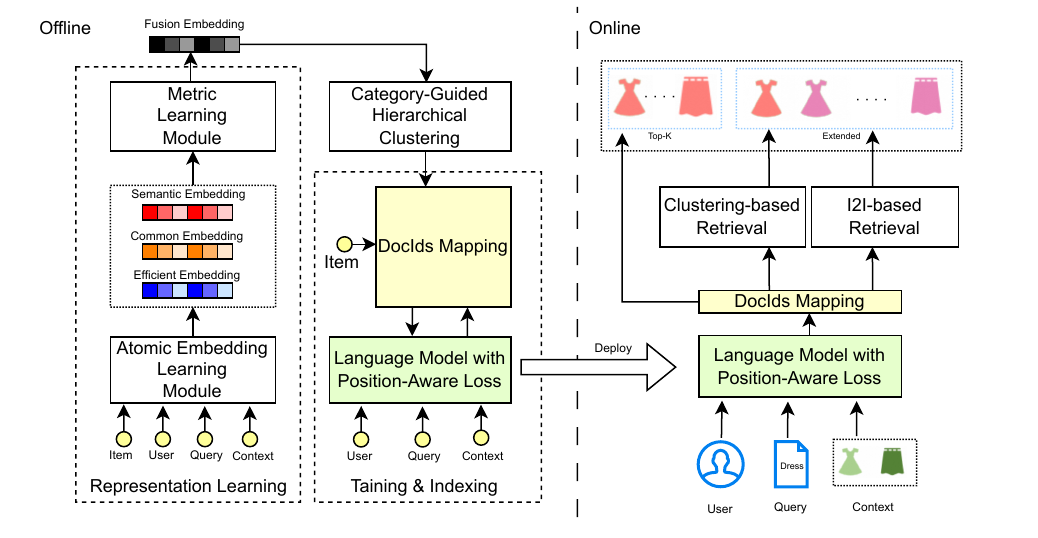}
    \caption{Overview of the proposed \textbf{Hi-Gen}. In the figure, the arrow indicates the data flow from input to output.
    }
    \label{fig_overflow}
\end{figure*}

Following the publication of Transformer Memory as a Differentiable Search Index(DSI)~\cite{tay2022transformer}, a significant number of GR approaches have been proposed~\cite{semanticenhance,salemi2023lamp,zhou2022ultron,hua2023index,yu2022generate,bevilacqua2022autoregressive,rajput2023recommender,wang2022neural,zhao2022dense,mehta2022dsi++,zhuang2022bridging,zhou2023dynamicretriever,kishore2023incdsi,raffel2020exploring,semanticenhance}. The performance improvement of General Relativity can be attributed to three primary factors: 1) refining the document representation; 2) enriching the docID encoding information; and 3) improving the language model used.


\textbf{Document Representaion}. Wang \etal ~\cite{wang2022neural} introduced an additional query generation process for each document based on the content. Then they bound the information of document content through training the Seq2Seq model with generated queries and their corresponding document identifiers to make the identifiers aware of the document semantics. In~\cite{rajput2023recommender}, a pre-trained content encoder was adopted to generate a semantic embedding for each item which has associated content features that capture useful semantic information. Lian \etal ~\cite{lian2023personaltm} utilized a transformer-based encoder-decoder network to jointly learn query embedding, document embedding, and personal embedding for learning the relationship between the query and docID, document and docID, as well as the query and the personalized information. However, there are relatively few works that explicitly learn discriminative representation to encode both the semantic and efficiency information about the item.

\textbf{DocID Generation}. Generally, a docID can be expressed as a single token (or atomic integers) or as a series of tokens, each of which could be a semantic numeric string or any arbitrary string. Tay \etal~\cite{tay2022transformer} investigated three kinds of docID generation strategies (\ie unstructured atomic identifiers, naively structured string identifiers, and semantically structured identifiers) and claimed that semantically structured identifiers lead to better indexing and retrieval capabilities for DSI. GenNet~\cite{sun2023learning} learns to tokenize documents into semantic docIDs via a discrete auto-encoding approach. TIGER~\cite{rajput2023recommender} introduces a hierarchical quantizer (RQ-VAE) that generates semantic ID representations for items. SE-DSI~\cite{semanticenhance} takes the generated pseudo query with explicit semantic meaning, called Elaborative Description, as the docID. Nevertheless, these methods solely focus on the semantic information of the documents but ignore the efficiency information. To capture both the semantic and efficiency information of items, we develop a discriminative representation process and then employ category-guided hierarchical clustering to generate docIDs.

\textbf{Language Model Optimization}. To train the language model in GR, NCI~\cite{wang2022neural} proposes a PAWA loss to tackle the significance variation problem when the same token appears in different places. PersonalTM~\cite{lian2023personaltm} utilizes hierarchical loss to increase the impact of topper level clusters. Though they have acknowledged the specificity of the docID, the inherent semantic and efficiency information buried in different positions of the docID remains uncovered. Therefore, position-aware loss is suggested to distinguish the importance of different positions and discover the inherent interrelation between different tokens within the position. Since recent results suggest that some properties of large language models emerge only for very large model sizes, some existing work has tried to improve the LM by increasing model parameters or using better models~\cite{tay2022transformer}.

\section{Proposed Methodology}
\sloppy{}

In this section, we introduce the proposed Hi-Gen model in detail. First, an overview of the model is illustrated. Subsequently, we clarify the procedures for discriminative representation learning. After that, the document identifier generation method is given. Then we introduce how to train the language model with position-aware loss. Finally, two variations of Hi-Gen are proposed to support real-time large-scale recall.

\begin{figure*}
\centering
\includegraphics[width=\linewidth]{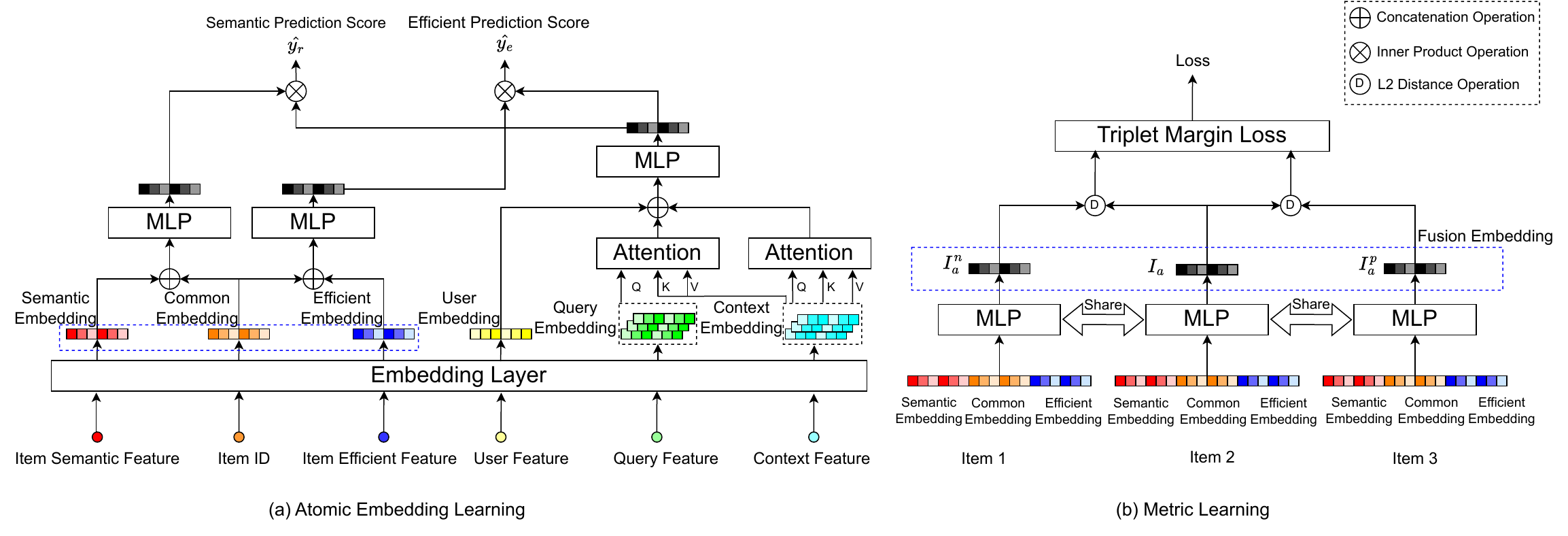}
    \caption{Overview of discriminative representation learning. }
\label{fig_metric_learning}
\end{figure*}
	
\subsection{Overview}
Figure \ref{fig_overflow} depicts the overall architecture of Hi-Gen. Hi-Gen consists of offline and online stages. At the offline stage, a discriminative representation learning model is first employed to learn a fusion embedding for each item. Then the learned representation is encoded into a semantically and efficiently structured document identifier (docID) through a category-guided hierarchical clustering process. Subsequently, a pre-trained language model (LM) equipped with a position-aware loss is fine-tuned using the training data. The LM is fed with the user, query, and context-related features and learns to output the corresponding item docIDs. At the online stage, the fine-tuned LM model generates the most related top-K docIDs with a given query. Two variants of Hi-Gen (\ie Hi-Gen-I2I and Hi-Gen-Cluster) are designed to rapidly expand results based on top-K decoded docIDs to support online large-scale recall. In what follows, we provide detailed descriptions of the proposed Hi-Gen.

\subsection{Discriminative Representation Learning}

Given a query, the E-commerce search system makes personalized product recommendations based on the user's historical shopping behaviors and preferences. In our work, we construct an E-commerce search algorithm model by taking into account four types of features: user, query, item, and context features. In particular, \textit{query features} are extracted from both the original query and its corresponding bigrams. \textit{Item features} consist of item ID, title, and some other statistical information. \textit{User features} consist of user ID, tags, and some other user-related information. \textit{Context features} include several user historical behavior sequences under a specified query, such as historical click and pay item sequences.

Using dense document representation to generate docID has been proven to be an effective method~\cite{tay2022transformer} in GR. However, the efficiency information about the item has never been taken into account in existing works. To eliminate this problem, we designed a discriminative representation learning model that captures both semantic relevance and efficiency information about the item. The model consists of two modules: 1) Atomic Embedding Learning Module, and 2) Metric Learning Module.


\subsubsection{Atomic Embedding Learning Module}
With user, query, item, and context features as input, three atomic embeddings (semantic, common, and efficient) are produced in this module, which represent semantic, universal and efficiency information of the item, respectively. Figure \ref{fig_metric_learning} (a) illustrates the architecture of the atomic embedding learning module. For simplicity, we represent the output of the embedding layer for the user feature, query feature, and context feature in a batch as $\textbf{x}_u \in \mathbb{R}^{B \times d_u}$, $ \textbf{x}_q \in \mathbb{R}^{B \times d_q \times d_k}$, and $\textbf{x}_c \in \mathbb{R}^{B \times d_c \times d_k}$, where $d_u$ and $d_k$ are embedding dimensions of each feature, and $d_q$ and $d_c$ are the sequence lengths for query and context, $B$ represents the batch size in training and inference stages.

To capture the underlying factors that explain variations in item data, we explicitly group the original features of an item into three parts: 1) \textit{item semantic feature}: the semantic information about an item, such as its title; 2) \textit{item ID}: unique identifier for an item; 3) \textit{item efficient feature}: some efficiency-related statistical indicators, such as click and pay count within a week. After embedding and feature dimension reduction processing, these three parts of features are converted into three low-dimensional atomic features: $\mathbf{x}_{i_s} \in \mathbb{R}^{B\times d_{i_{s}} }$, $\mathbf{x}_{i_c} \in \mathbb{R}^{B \times d_{i_{c}}}$ and $\mathbf{x}_{i_e} \in \mathbb{R}^{B\times d_{i_{e}}}$, which are named as semantic embedding, common embedding, and efficient embedding, respectively. Notably, these three atomic item embeddings could have the same feature dimension (\ie $d_{i_s}=d_{i_c}=d_{i_e}$). To capture both semantic and efficiency information of items, a multi-task two-tower embedding model~\cite{yu2021dual} that consists of a user tower network and an item tower network is employed in which relevance prediction and click prediction tasks are deployed as downstream tasks as shown in Fig~\ref{fig_metric_learning}. 

In the user tower network, we utilize two attention modules to make the model learn and focus on the important information within user and query embeddings: UserSelfAttention and UserQueryAttention. In practice, we compute the attention function on a set of queries simultaneously, packed together into a matrix $\mathbf{Q}$. The keys and values are also packed together into matrices $\mathbf{K}$ and $\mathbf{V}$. Thus we computerize the matrix of outputs as:
\begin{equation}
    \begin{aligned}
        \mathbb{Z}(\textbf{Q}, \textbf{K}, \textbf{V}) = reshape(softmax(\frac{\textbf{Q} \cdot \textbf{K}^T}{\sqrt{d_k}} ) \cdot \textbf{V})
    \end{aligned}
    \label{eq:user_attention}
\end{equation}

Therefore, the UserSelfAttention and UserQueryAttention can be represented as $\mathbf{Z}(\mathbf{x}_c, \mathbf{x}_c, \mathbf{x}_c) $ and $\mathbf{Z}(\mathbf{x}_q, \mathbf{x}_c, \mathbf{x}_c)$. The dimensions of UserSelfAttention and UserQueryAttention are $\mathbb{R}^{B \times (d_c \times d_k)}$ and $\mathbb{R}^{B \times (d_q \times d_k)}$, respectively. The concatenation of the two attentions and $\textbf{x}_u$ is fed into a multilayer perceptron (MLP). Thus, the hidden layer $\mathbf{U}$ of the user tower network with dimension $\mathbb{R}^{B \times d_e}$ can be expressed as follows:
\begin{equation}
    \begin{aligned}
\textbf{U} = MLP(concat(\textbf{x}_u, \mathbb{Z}(\textbf{x}_c, \textbf{x}_c, \textbf{x}_c), \mathbb{Z}(\textbf{x}_q, \textbf{x}_c, \textbf{x}_c))).
    \end{aligned}
    \label{eq:user_MLP}
\end{equation}



As for the item tower network, two MLPs are employed for relevance prediction and click prediction tasks. The common embedding of the item is shared between these two tasks. The outputs of these two tasks are formulated as follows:

\begin{equation}
    \begin{aligned}
           \widehat{y}_r &= \mathcal{N}(\mathbf{U}) \odot \mathcal{N}(MLP(concat(\mathbf{x}_{i_s}, \mathbf{x}_{i_c})))\\
            \widehat{y}_c &= \mathcal{N}(\mathbf{U}) \odot \mathcal{N}(MLP(concat(\mathbf{x}_{i_e}, \mathbf{x}_{i_c})))
    \end{aligned}
    \label{eq:prediction_score}
\end{equation}
where $\odot$ is the dot multiplication operation and $\mathcal{N}(\cdot)$ is the $L2$-norm normalization function.

To train the atomic embedding learning module, we optimize the loss function as follows:
\begin{equation}
    \begin{aligned}
    \mathcal{L}_{1} &= loss(y_r, \widehat{y}_r) + w_c \times loss(y_c, \widehat{y}_c)
\end{aligned}
    \label{eq:embedding_loss}
\end{equation}
where $y_r$ is the relevance label and $y_c$ is the click label of the item. $w_c$ is a hyperparameter that is used to balance the relevance task and the click task in training. $loss(\cdot)$ can be any suitable loss function. In this paper, we employ the cross-entropy loss function for Eq.\ref{eq:embedding_loss}.




\subsubsection{Metric Learning Module} Through the atomic embedding learning module, we acquire three atomic embeddings: $\mathbf{x}_{i_s}$, $\mathbf{x}_{i_c}$, and $\mathbf{x}_{i_e}$.
However, since the two-tower network mainly focuses on modeling the similarity between item and user, the generated embeddings may be weak at distinguishing the semantic and efficiency information of similar items. As we know, metric learning can map the original features into some appropriate feature space where similar items are closer to each other, and dissimilar items are further apart~\cite{hoffer2015deep,kaya2019deep,yang2006distance,kulis2013metric}. Thus, it is a suitable choice to introduce metric learning to refine the learned atomic embedding features.


Figure \ref{fig_metric_learning} (b) illustrates the architecture of our metric learning module. We leverage an MLP to learn a compacted yet discriminative feature embedding from the combination of efficient, semantic, and common embeddings of the item as shown in Eq.\ref{eq:fusion_embedding}. And, we denote the refined fusion embedding as $\textbf{I}_a \in \mathbb{R}^{d}$, where $d$ is the feature dimension.

\begin{equation}
    \textbf{I}_a = MLP(concat(\mathbf{x}_{i_c}, \mathbf{x}_{i_e}, \mathbf{x}_{i_s}))
    \label{eq:fusion_embedding}
\end{equation}


Subsequently, a triplet margin loss is used to reduce the distance between items with the same user behaviors (\eg click or pay) on the same exposure page view (PV) and increase the distance between items with different behaviors. We collect a large amount of online real-exposure page information to form a training data set $PV = \{PV_{1}, PV_{2}, \dots, PV_{n} \}$ where $PV_{i} = \{(I_{a}^{j}, y_{a}^{j}) | j \in (1, 2, \dots, n_{i})\}$ and construct the following optimization function:

\begin{equation}
\begin{split}
   &\mathcal{L}_{2} = \sum_{i=0}^{n}\sum_{\mathbf{I}_{a}^{j}}\sum_{\mathbf{I}_{a}^{j_n}} \sum_{\mathbf{I}_{a}^{j_p}}max\left\{m, dis(\mathbf{I}_{a}^{j}, \mathbf{I}_{a}^{j_n}, \mathbf{I}_{a}^{j_p})\right\} \\
   &dis(\mathbf{I}_{a}^{j}, \mathbf{I}_{a}^{j_n}, \mathbf{I}_{a}^{j_p}) =  dist(\mathbf{I}_{a}^{j}, \mathbf{I}_{a}^{j_n}) - dist(\mathbf{I}_{a}^{j}, \mathbf{I}_{a}^{j_p}) \\
    s.t. \quad {\mathbf{I}_{a}^{j_n}} &= \{\mathbf{I}_{a}^{k} \in PV_{i} \mid y_{a}^{k} \neq y_{a}^{j}\} \quad \forall i, k, \mathbf{I}_{a}^{j}\\
        \mathbf{I}_{a}^{j_p} &= \{\mathbf{I}_{a}^{k} \in PV_{i} \mid y_{a}^{k} = y_{a}^{j}\} \quad \forall i, k, \mathbf{I}_{a}^{j}\\
\end{split}
\label{eq:triplet_loss} 
\end{equation}
where $m$ is a positive hyper-parameter and $dist(\cdot)$ is a distance function. $y_{a}^{j}$ stands for the click label of the j-th sample in the PV, $j_p$ stands for the sample with the same click label as the j-th sample, and $j_n$ stands for the sample with different click label as the j-th sample. In this paper, we use the L2-norm distance as the $dist$ function.

\subsection{Document Identifiers (docIDs) Generation}
Items in a large-scale E-commerce search engine are naturally organized and stored according to a specific hierarchical category tree structure. For instance, a red long dress can be categorized as (\textit{clothing}, \textit{dress}) where \textit{clothing} is the name of the upper category and \textit{dress} is the name of the lower leaf category. Utilizing the hierarchical category tree aids in more efficient item retrieval because it brings in useful semantic information.

Using hierarchical clustering, the docID of an item can be acquired once the dense vector representation is given~\cite{tay2022transformer}. However, this method yields an unintelligible docID that is devoid of real category information. To solve this problem, we designed a category-guided hierarchical clustering method for document identifier generation, and thus the generated docID has a strong and clear semantic meaning.


\begin{figure}
    \centering
    \includegraphics[width=\linewidth]{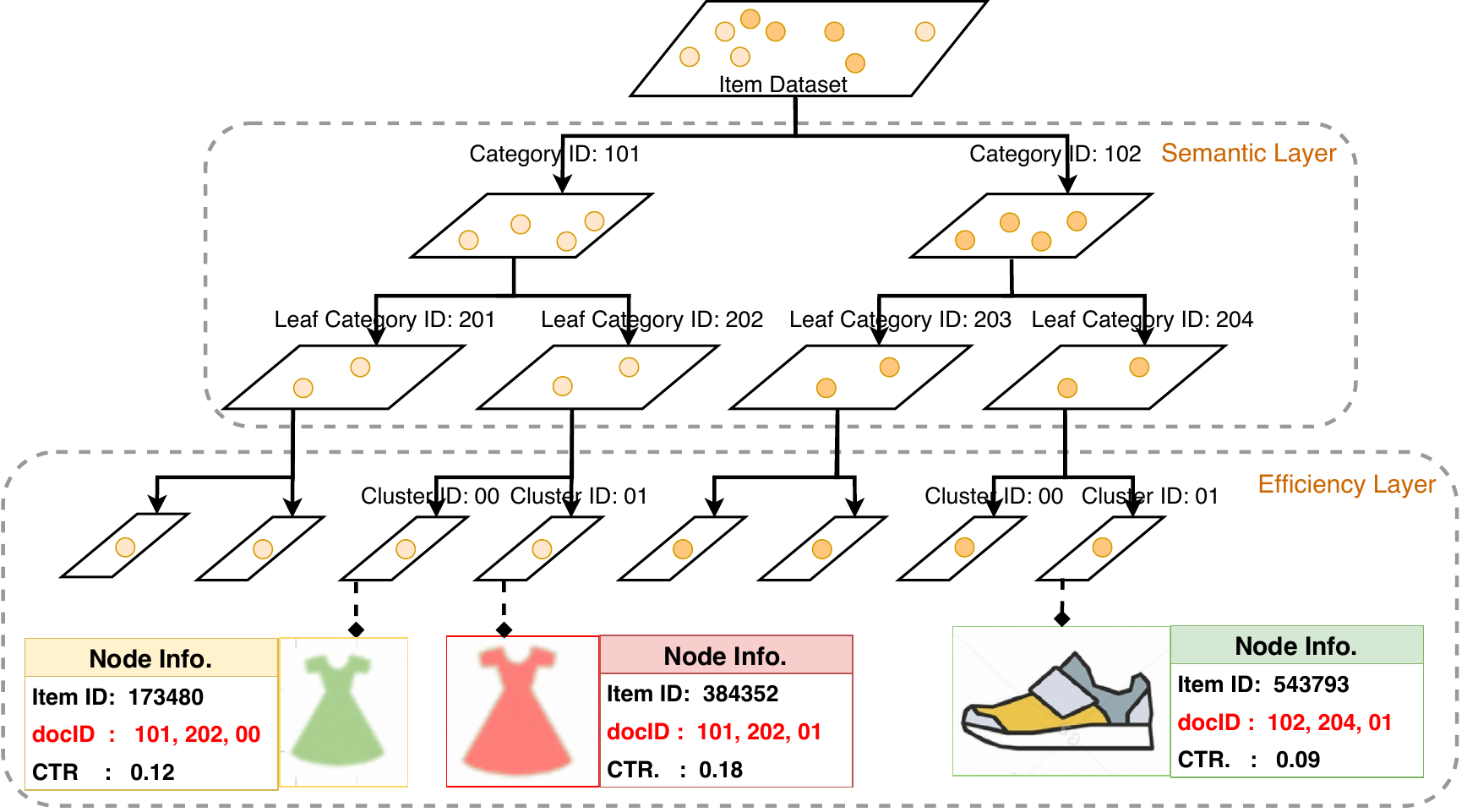}
    \caption{An illustration of the category-guided hierarchical clustering method that produced the docID.}
    \label{fig_category_hierarchical_clustering}
\end{figure}

As shown in Figure \ref{fig_category_hierarchical_clustering}, we first classify all items according to the category tree. Items that belong to the same leaf category undergo hierarchical clustering. Each category has its distinct category ID. For example, dress can be assigned to a leaf category ID 202. In addition, we calculate an efficient score (\eg click-through rate (CTR) score) for each item to calculate the average efficient score for each cluster in the efficiency layer. The results of an item encoded by the efficiency layer are appended to the end of its category path to produce the final docID. The proposed category-guided hierarchical clustering algorithm is detailed in Algorithm \ref{alg:category-guided hierarchical clustering}.

  \begin{algorithm}
       \normalem
  \SetAlgoLined
  \SetKwData{Left}{left}\SetKwData{This}{this}\SetKwData{Up}{up}
  \SetKwFunction{Union}{Union}\SetKwFunction{categoryHierarchicalCluster}{CategoryGuidedHierarchicalCluster}
  \SetKwFunction{Union}{Union}\SetKwFunction{hierarchicalCluster}{HierarchicalKmeansCluster}
  \SetKwFunction{Union}{Union}\SetKwFunction{kmeansCluster}{KmeansCluster}
  \SetKwInOut{Input}{input}\SetKwInOut{Output}{output}
  \SetKwProg{Fn}{Function}{:}{}

  \Input{Items' embedding: $\mathbf{I}$;\\ 
  Items' efficient scores: $ES$;\\
  Category paths in the semantic layer: $P$;\\
  Max length of the docID: $L$;\\
  Number of cluster: $K$;\\
  Max cluster size: $CS$;\\
  Number of Categories: $C$; \\
  Number of Item: $N$; 
  }
  \Output{The docIDs: $D$;\\ efficient scores for each cluster: $E$;}

  \Fn{\categoryHierarchicalCluster{$\mathbf{I}$, $ES$, $P$, $L$, $K$, $CS$, $C$, $N$}}{
    \For {each $p \in P_{1:C}$} {
        \emph{ $D_{current}$ $\leftarrow$ [p] }\;
       \tcp{Clustering for all items under the category path:p}
	\emph{ $KC_{1:K}$ $\leftarrow$ \kmeansCluster($\mathbf{I}^p$, $K$) }\; 
        \For{$i \leftarrow 1$ \KwTo $K$} {
            \If{$|KC_{i}| >$ CS} {
                \emph{$D_{rest}\leftarrow$ \hierarchicalCluster($\mathbf{I}_{i}^p$, $K$, $L$)} \;
            }
            \Else {
                \emph{$D_{rest} \leftarrow [0, ...., |CL_{i}| -1]$}\;
                
            }
            \emph{$E_{cluster}$ $\leftarrow$ $avg(ES_{i}^p)$}\;
            \emph{$D_{cluster} \leftarrow$ ($D_{current} , i, D_{rest}$})\;
	    \emph{D $\leftarrow D\cup D_{cluster}$}\;
            \emph{E $\leftarrow E\cup E_{cluster}$}\;
        }
    }
  }
 
  \caption{Generating the docID by category-guided hierarchical clustering.}
  \label{alg:category-guided hierarchical clustering}
\end{algorithm}


\subsection{Language Model With Position-aware Loss }

Our unique docIDs include three distinguishing characteristics: 1) Prediction errors on the first several digits of the docID will be more significant than those on subsequent digits; 2) Improved relevance results from choosing to predict a category at the semantic layer that is pertinent to the target category rather than one that is irrelevant; 3) Predicating a cluster with a similar efficient score to the target cluster at the efficiency layer will lessen the efficiency divergence. To address these issues, a position-aware loss is proposed.

        


As shown in Eq.\ref{eq:adaptive_loss}, the position weight in the $t$ step (\ie $\textbf{w}_t^p$) consists of the positional hierarchical, semantic, and efficient weights, which are represented as $\textbf{w}_t^h$, $\textbf{w}_t^s$, and $\textbf{w}_t^e$, respectively. Greater weight is given to the earlier token by $\textbf{w}_t^h$. $\textbf{w}_t^s$ causes a greater loss at the semantic layer for the irrelevant category. At the efficiency layer, $\textbf{w}_t^e$ attempts to generate tokens that are as efficient as the target token. $\lambda_h$, $\lambda_s$, and $\lambda_e$ are hyperparameters that help to balance the effect of positional hierarchical, semantic, and efficient weight. The semantic layer's length is $S$, while the docID's length is $L$. The loss function for each token in the docID is represented by $loss$. $dist(\cdot)$ is a distance function that judges how similar clusters are in efficiency, while $relevance$ determines whether two categories are semantically similar or not. $E$ represents the efficient score which is calculated in Algorithm \ref{alg:category-guided hierarchical clustering}. In our work, $dist(\cdot)$ is defined as the difference of the Click-Through Rate (CTR), and bert-sentence-similarity\footnote{https://www.kaggle.com/code/eriknovak/pytorch-bert-sentence-similarity} is utilized as $relevance$. $\textbf{w}_t^h$ is a decay function, we set it to $\frac{e^{L-t}}{\sum_{i=0}^{L}e^i}$. In addition, we use the cross-entropy loss as $loss(\cdot)$.

\begin{equation}
    \begin{aligned}
    \mathcal{L}_{3} &= \sum_{t=0}^{L}\textbf{w}_t^{p}loss(y_t | \textbf{y}_{<t}, \textbf{x}_u, \textbf{x}_q, \textbf{x}_h, \widehat{y}_t; \theta) \\
            s.t. \quad \textbf{w}_t^p &= \lambda_h \times \textbf{w}_t^h
            + \lambda_s \times \textbf{w}_t^s
            + \lambda_e \times \textbf{w}_t^e \\
        \textbf{w}_t^s &=
        \begin{cases}
        1, & \text{relevance($y_{t}$, $\widehat{y}_{t}$)=0 and t $\leq$ S} \\
        0, & otherwise
        \end{cases}\\
        \textbf{w}_t^e &=  \begin{cases}
        dist(E_{y_{t}}, E_{ \widehat{y}_{t}}) , & \text{t $\textgreater$ S} \\
        0, & otherwise
        \end{cases} \\
     \sum_{t=0}^{L}\textbf{w}_t^{h} &= 1, \textbf{w}_t^{h} > \textbf{w}_{t+1}^{h}, \mathbf{w}_{t}^{h} > 0 \quad \forall t. \\
    \end{aligned}
    \label{eq:adaptive_loss} 
\end{equation}

\subsection{Online Serving Optimization}
In a realistic large-score personalized search engine, retrieval performance is influenced by both the quantity and quality of recalled items. However, the language model is unable to generate numerous docIDs due to high latency in inference. To solve this issue, we propose two variants of Hi-Gen (\ie Hi-Gen-I2I and Hi-Gen-Cluster) for efficient online large-scale recall.

I2I retrieval~\cite{i2i_1167344,Yang2020LargeSP,i2iretrieval2,Qian630079,linden2003} uses the consumer's recent behaviors as triggers to retrieve recommended items. Hi-Gen-I2I treats the limited items that are generated from Hi-Gen as triggers and uses an I2I retrieval method (\eg Swing\footnote{Swing is an i2i algorithm proposed by Alibaba Group and widely used in the industry.}) to perform the recall process. Thus, it also can expand recall results quickly.

With category-guided hierarchical clustering, extended results can be produced efficiently and rapidly because items in the same cluster share similarities. All items in the same cluster are taken into consideration by the Hi-Gen-Cluster. Namely, Hi-Gen-Cluster-K regards items that share the same top K tokens in their docIDs produced by the language model as the expanded recalled items.

\section{Experiments Setup}
 

\subsection{Datasets} 
We conduct experiments on two datasets (\ie \textbf{AOL4PS}~\cite{aol4ps} and \textbf{AEDST\footnote{Our code and dataset will be publicly available soon.}}) to evaluate Hi-Gen for large-scale personalized search. Below, we describe the details of these two datasets.


\textbf{AOL4PS} is an open-source large-scale dataset tailored for personalized search research, that contains the user, the query, the corresponding clicked documents (URLs), and timestamps indicating the timeline of each user's interaction history. Table~\ref{tab:dataset} illustrates the respective quantities in both the training and testing sets, detailing the number of samples, users, queries, and documents in each. To meet the input requirements of Hi-Gen, we extract the first two words of each URL (\eg www.spiritplay.org represented as www.spiritplay) as a category and construct a one-layer hierarchical category tree. Furthermore, we observed a 65$\%$ overlap between the training and testing sets in this dataset. To assess the generalization capability of the Hi-Gen model, we excluded samples from the testing set that had already appeared in the training set for zero-shot experimental analysis. The specific approach aligns with the methodology employed by PersonalTM~\cite{lian2023personaltm}. Since AOL4PS lacks relevance labels, we apply the relevance model, bert-sentence-similarity\footnote{https://www.kaggle.com/code/eriknovak/pytorch-bert-sentence-similarity}, to evaluate the relationship between the query and documents in AOL4PS. The relevance label is 1 if the relevance score is more than 0.5, and 0 otherwise.

 
\textbf{AEDST} is a billion-scale industrial dataset sourced from AliExpress, a real large-scale E-commerce search system, including the user, the query, the corresponding items, the click and relevance label. Organized in chronological order of user behavior, we designate the samples from the first 7 days as the training set and the samples from the subsequent day as the testing set. The detailed distribution of quantities is shown in Table \ref{tab:dataset}. A detailed analysis revealed that 87.33$\%$ of the samples in the testing set had never appeared in the training set. This phenomenon is attributed to the nature of large-scale E-commerce search systems, where users typically do not click or purchase the same item under the same query in a short period. Hence, the testing set of AEDST dataset can be directly utilized to evaluate the model's generalization performance.

\begin{table}[htbp]
    \caption{The detailed information of different datasets.}
    \label{tab:dataset}
    \begin{center}
        \renewcommand\arraystretch{1.5}
        \begin{tabular}{cccccc}
        \hline
        \multicolumn{2}{c}{Type} & $\#$Sample & $\#$Query &$\#$User & $\#$Doc\\
        \hline
        \multirow{2}{*}{AOL4PS} & Training & 218,559 & 82,942 & 12,907 & 54,952 \\
        & Testing  & 53,357 & 26,584 & 12,907 & 20,327 \\
        \hline
        \multirow{2}{*}{AEDST} & Training  & 1687M & 43M & 27M & 21M \\
        & Testing  & 24M & 0.4M & 0.2M & 5M \\
             
        \hline
           
    \end{tabular}
    \end{center}

\end{table}        
	
\subsection{Baselines}


To verify Hi-Gen's effectiveness, we compare it with three existing retrieval methods: sparse retrieval, dense retrieval and generative retrieval on the AOL4PS dataset.

\textbf{Sparse Retrieval}. We chose  BM25\footnote{https://github.com/dorianbrown/rank\underline{~}bm25} as the baseline of sparse retrieval model. BM25 is a widely-used sparse retrieval in information retrieval that retrieve documents based on their relevance to a query, factoring in term frequency and document length.

\textbf{Dense Retrieval}. Classic dense retrieval models, including P-Click\cite{pclick}, SLTB\cite{sltb}, HRNN\cite{Ge2018PersonalizingSR}, GRADP\cite{zhou2019dynamic}, are chosen to shows that Hi-Gen shows better performance than traditional \textit{retrieve-then-rank} models. P-Click calculates the personalized score of a document as the percentage of clicks on this document from past queries that are identical to the input query. SLTB
outputs personalized recalled items by utilizing diverse clicked-based or topic-based features. HRNN extends previous RNN-based session modeling with one additional GRU level that models the user activity across sessions and provides a seamless way of transferring the knowledge acquired on the long-term dynamics of the user interest to session-level. GRADP uses RNN-based networks to extract short-term and long-term user profiles from personalized historical information.

\textbf{Generative Retrieval}. In order to prove that Hi-Gen is an effective generative retrieval with better metrics, we compare it with the classic generative retrieval model DSI~\cite{tay2022transformer} and the state-of-the-art model PersonalTM~\cite{lian2023personaltm}. DSI is a \textit{sequence-to-sequence} generative retrieval model outperforms strong baselines by directly generating relevant document
identifiers from queries without relying on an explicit index. It is a new information retrieval paradigm aims to address the limitations of dense retrieval framework based on the similarity score.  PersonalTM proposes a novel decoder architecture that effectively leverages personalized context without increasing trainable parameters, and applys a hierarchical loss function to optimize the performance. 

In the AEDST, we compare Hi-Gen with BM25, P-Click and DSI. We use T5-Small\footnote{https://huggingface.co/t5-small} and T5-Base\footnote{https://huggingface.co/t5-base} as the transformer backbones for DSI and Hi-Gen in order to examine the effects of various parameter sizes.

	
\subsection{Implementation Details}
 
All experiments are conducted on the Tesla A100 platform using the Adam optimizer. The representation learning model, metric learning model\footnote{https://github.com/KevinMusgrave/pytorch-metric-learning}, and language model are assigned learning rates of 0.0001, 0.00001, and 0.00005, respectively. Correspondingly, their batch sizes are set at 512, 10, and 64. The representation and metric learning models yield embeddings with dimensions of 256 and 768, respectively. Fast-kmeans\footnote{https://pypi.org/project/fast-pytorch-kmeans/} in which k is 10 is utilized to produce the docID. In cases where the cluster size exceeds 100, the hierarchical clustering algorithm is executed recursively; otherwise, termination occurs. The embedding of dimension 768 is harnessed to generate the docID for all DSI baselines, originating from a pre-trained BERT~\cite{devlin2018bert}\footnote{https://huggingface.co/bert-base-uncased} model. For the proposed Hi-Gen, we utilized the following settings: (1) $w_{c}$: 1.0; (2) $m$: 0.1; (3) $\lambda_h$: 0.8; (4) $\lambda_s$: 0.1; (5) $\lambda_e$: 0.1; (6) $S$: 1. Predictions are obtained by decoding the language model using the beam search methodology. Additionally, all experiments adopt k-fold cross-validation, with k set to 3.
	
\subsection{Evaluation Metrics}
 
The retrieval accuracy of the model for k items is denoted as Recall@k. This metric serves as a benchmark for evaluating the performance of various retrieval algorithms across diverse datasets. Additionally, six supplementary metrics are utilized to evaluate the effectiveness of generative retrieval in online testing: RecallNum, click-through rate (CTR), conversion rate (CVR), PayCount, gross merchandise value (GMV), and Latency. RecallNum signifies the total number of items retrieved by the online search system. CTR and CVR represent simplified forms of click-through rate and conversion rate, respectively. PayCount quantifies the number of transaction orders completed by users after selecting a sponsored result. The abbreviated form of gross merchandise value is referred to as GMV. Latency serves as a metric to quantify the processing time for retrieval.
	
\section{Experimental Results}
\sloppy{}
	
\subsection{Main Results}
Table \ref{tab:main_result_aol4ps} illustrates the experimental outcomes for AOL4PS, providing compelling evidence that Hi-Gen surpasses other retrieval methods in large-scale personalized search systems. Hi-Gen beats the state-of-the-art model (PersonalTM) by 3.30$\%$ and 7.73$\%$ in Recall@1 and Recall@10, respectively. Especially, all generative retrievals including DSI outperform baselines of sparse and dense retrievals in Recall@1 and Recall@10.

\begin{table}[htbp]
    \caption{Performance comparison in AOL4PS. The metrics of other baseline methods are from ~\cite{lian2023personaltm}. Metrics with $^{*}$ represent the best results of baselines. $\Delta_p$ denotes the average Recall@k improvement over the best metric of baselines.}
    \label{tab:main_result_aol4ps}
    \begin{center}
    \renewcommand\arraystretch{1.5}
    \begin{tabular}{ccc}
        \hline
        Method & Recall@1 & Recall@10 \\
        \hline
        BM25 & 21.61 & 32.87 \\

        P-Click & 59.56 & -- \\
        SLTB & 71.09 & -- \\
        HRNN & 76.53 & -- \\
        GRADP & 77.06 & -- \\
        DSI & 67.42 & 75.84 \\
         PersonalTM & $79.60^{*}$ & $87.47^{*}$ \\
        \hline
        \textbf{Hi-Gen} & \textbf{82.90} & \textbf{95.20} \\
        $\Delta_p$ & +3.30 & +7.73 \\
        \hline
    \end{tabular}
    \end{center}
\end{table}

In order to evaluate the effectiveness of Hi-Gen in a larger and more realistic dataset, we compare it with classic models of sparse, dense and generative retrievals: BM25, P-Click and DSI. As indicated in Table \ref{tab:main_result_ae}, Hi-Gen outperforms the best baseline model (BM25) on Recall@1, Recall@50, and Recall@100, improving by 4.62$\%$, 25.87$\%$, and 28.37$\%$, respectively. Compared to T5-Small, T5-Base has twice as many model transformer layers. Based on the research metrics of DSI and Hi-Gen, we could conclude that using T5-Base as the backbone outperforms T5-Small. It suggests that better results are obtained with more layers. Surprisingly, in AEDST, P-Click and DSI perform even worse than BM25. The main reason is an abundance of zero-shot cases in AEDST. As seen in Table \ref{tab:zero_shot_aol4ps}, this phenomenon also occurs in zero-shot circumstances of AOL4PS.
	
\begin{table}[htbp]
    \caption{Performance evaluation in AEDST.}
    \label{tab:main_result_ae}
    \begin{center}
        
    \renewcommand\arraystretch{1.5}
    \begin{tabular}{cccl}
        \hline
         Method & Recall@1 & Recall@50 & Recall@100  \\
        \hline
         BM25 & $4.01^{*}$ & $29.26^{*}$ & $33.84^{*}$ \\
         P-Click & 0.23 & 6.20 & 9.99 \\ 
         DSI-T5-Small & 0.06 & 1.94 & 2.45\\
         DSI & 1.89 & 11.10 & 13.59 \\ 
        \hline
         Hi-Gen-T5-Small & 6.97 & 51.11 & 55.93  \\
         \textbf{Hi-Gen} & \textbf{8.63} & \textbf{55.13} & \textbf{62.21} \\
            $\Delta_p$ & +4.62 & +25.87 & +28.37\\
        \hline
    \end{tabular}
     \end{center}
\end{table}

\subsection{Ablation Study}	
To figure out the roles of different components in the model, we performed ablation research on Hi-Gen in AOL4PS. We offer four Hi-Gen variants: 1) w/o position-aware loss, where the loss is equal to DSI; 2) w/o category-guided clustering, in which the clustering technique is the same as DSI; 3) w/o metric learning, where metric learning is replaced with ordinary concatenation; 4) w/o discriminative representation, in which the embedding is identical to DSI.

In Table \ref{tab:ablation_study_ae}, the removal of any one of the four components of Hi-Gen results in a drop in performance, indicating the effectiveness of each component. Based on the results of experiments, we discover that category-guided hierarchical clustering is the most important part of Hi-Gen. When category-guided hierarchical clustering is eliminated, Recall@1 and Recall@10 drop by 21.22$\%$ and 22.71$\%$, respectively. Apart from category-guided hierarchical clustering, Hi-Gen also heavily relies on discriminative representation. Removing discriminative representation will result in a 17.00$\%$ and 18.30$\%$ decrease in Recall@1 and Recall@10, respectively. Furthermore, removing metric learning will decrease 8.02$\%$ and 8.86$\%$ in Recall@1 and Recall@10 in Hi-Gen. It proves that using discriminative representations learned through representation and metric learning, rather than BERT embedding, leads to significantly superior results for Hi-Gen. What's more, Hi-Gen will reduce Recall@1 and Recall@10 by 6.51$\%$ and 7.16$\%$, respectively, when position-aware loss is not utilized.

	\begin{table}[htbp]
		\caption{The results of ablation study in AOL4PS.}
		\label{tab:ablation_study_ae}
            \begin{center}
            \renewcommand\arraystretch{1.5}
		\begin{tabular}{ccl}
			\hline
			Method & Recall@1 & Recall@10 \\
			\hline
			\textbf{Hi-Gen} & \textbf{82.90} & \textbf{95.20} \\
			 w/o position-aware loss & 76.39 & 88.04 \\ 
			w/o category-guided clustering & 61.68 & 72.49 \\
			w/o metric learning & 74.88 & 86.34  \\
			 w/o discriminative representation & 65.90 & 76.90 \\
			\hline
		\end{tabular}
    
            \end{center}
	\end{table}

\subsection{Zero-Shot Analysis}
AEDST is a very sparse data set because zero-shot samples make up about 87.33$\%$ of the test data samples. The results of the experiment are shown in Table \ref{tab:main_result_ae}, where DSI performs far worse than Hi-Gen or even BM25. This result is in line with the findings published in ~\cite{lian2023personaltm}. We conducted experiments on the AOL4PS zero-shot scenarios to better illustrate Hi-Gen's effectiveness in such a scenario. Amazingly, Hi-Gen outperforms all baselines on any metric, as Table \ref{tab:zero_shot_aol4ps} illustrates. Moreover, compared to DSI, Hi-Gen yields notable gains in Recall@1 and Recall@10 of 36.78$\%$ and 41.07$\%$, respectively.
	
\begin{table}[htbp]
    \caption{The results of zero shot cases in AOL4PS. The metrics of baseline methods are from ~\cite{lian2023personaltm}.}
    \label{tab:zero_shot_aol4ps}
    \begin{center}
        
    \renewcommand\arraystretch{1.5}
    \begin{tabular}{cccl}
        \hline
        Method & Recall@1 & Recall@10 \\
        \hline
         BM25 & 20.97 & 31.03 \\
         DSI & 8.14 & 16.76 \\
         PersonalTM & $37.19^{*}$ & $56.94^{*}$ \\
         \hline
         \textbf{Hi-Gen} & \textbf{44.92} & \textbf{57.83} \\
          $\Delta_p$ & +7.73 & +0.89\\
        \hline
    \end{tabular}
    \end{center}
\end{table}


\subsection{The Impact Of Hi-Gen Variants}

\begin{figure}[htbp]
\centering
\includegraphics[width=\linewidth]{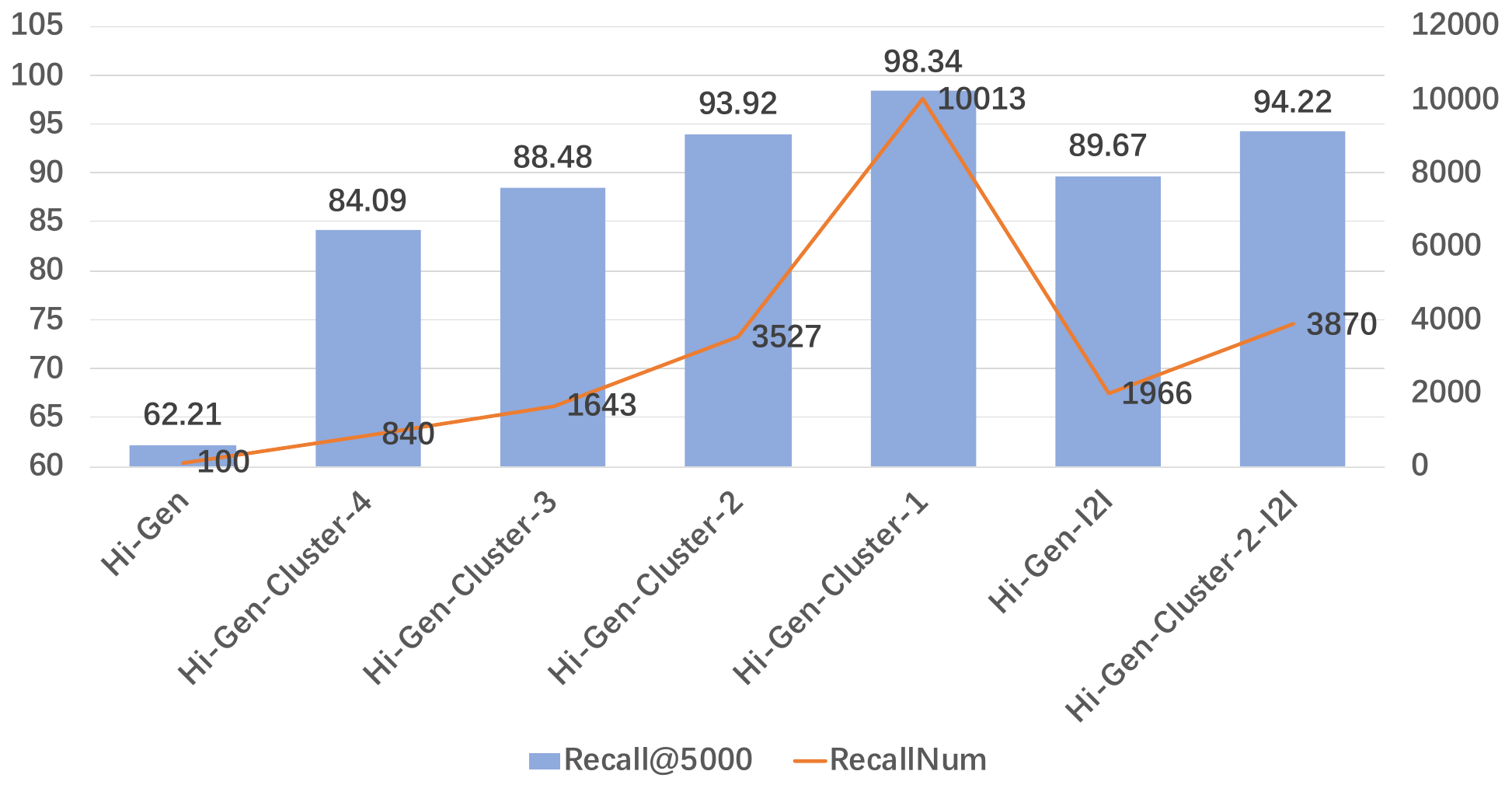}

\caption{Recall@5000 and RecallNum for different variants of Hi-Gen in AEDST.}
\label{fig:hi_gen_varients}
\end{figure}
 
    In the online serving phase, we propose two variants of Hi-Gen: Hi-Gen-Cluster and Hi-Gen-I2I. All products that have the same top K tokens as the 100 docIDs generated by the language model are designated as recalled items by Hi-Gen-Cluster-K. Figure \ref{fig:hi_gen_varients} illustrates how Recall@5000 and RecallNum of the Hi-Gen-Cluster increase as K decreases. Though Hi-Gen-Cluster-1 has improved recall@5000 by 36.13$\%$ compared to Hi-Gen, it recalls too many items, which is unbearable in the online system. Compared to Hi-Gen-Cluster-1, combining Hi-Gen-Cluster-2 and Hi-Gen-I2I (Hi-Gen-Cluster-2-I2I) is better since its RecallNum is reduced by 61.35$\%$, but Recall@5000 is only reduced by 4.12$\%$. Hi-Gen-Cluster-2-I2I has improved Recall@5000 by 0.3$\%$, 4.55$\%$, and 32.01$\%$, respectively, in contrast with Hi-Gen-Cluster-2, Hi-Gen-I2I, and Hi-Gen.

\subsection{Online Experiments}
	To examine Hi-Gen's performance in a realistic E-commerce search system, we implemented it in a real-word large-scale online E-commerce search engine. To ensure accuracy, Hi-Gen-Cluster-2-I2I is utilized to produce a sufficient number of recall items. A live experiment is conducted for a week using an A/B testing framework. In the control group, 4.0$\%$ of users are chosen at random and shown search results that don't come from Hi-Gen. An extra 4.0$\%$ users are chosen at random in the experiment group, and shown search results contain items produced by Hi-Gen. As indicated in Table \ref{tab:online_metrics}, Hi-Gen increased RecallNum, CTR, CVR, PayCount, and GMV by 6.89$\%$, 0.83$\%$, 1.51$\%$, 0.92$\%$, and 1.42$\%$, respectively. It implies that Hi-Gen has significant advantages in terms of retrieving efficient items in a practical large-scale E-commerce search engine. Furthermore, Hi-GEN can meet online needs without increasing latency when compared to current complex online retrievals because it just generates a limited number of items by the language model.

	\begin{table}[htbp]
		\caption{Online A/B experimental results.}
		\label{tab:online_metrics}
            \begin{center}
            \renewcommand\arraystretch{1.5}
		\begin{tabular}{cccccl}
			\hline
			RecallNum & CTR & CVR  & PayCount & GMV & Latency \\
			\hline
			+6.89$\%$ & +0.83$\%$& +1.51$\%$ & +0.92$\%$ & +1.42$\%$ & +0.56$\%$\\
			\hline
		\end{tabular}
            \end{center}
	\end{table}

	\section{Conclusion}
	In this work, we propose Hi-Gen, a generative retrieval for large-scale personalized E-commerce search systems. Unlike other generative retrievals, Hi-Gen encodes efficient and semantic information during docID generation meanwhile concurrently making full use of positional information during the decoding stage. In the docIDs generation phase, a representation learning model along with metric learning is designed to learn discriminative feature representations of products to capture both
semantic relevance and efficiency information. Then, category-guided hierarchical clustering is proposed to enhance the semantics of the docID. Position-aware loss is suggested to imitate positional information during the decoding stage. Based on the experiment's results, Hi-Gen is capable of achieving state-of-the-art performance in open-source and industrial large-scale personalized search datasets. Furthermore, we utilize Hi-Gen in a practical, extensive E-commerce search system and demonstrate its efficacy via online A/B testing. Notably, Hi-Gen performs admirably at resolving zero-shot scenarios.

\vspace{12pt}

\normalem
\bibliographystyle{IEEEtran}
\bibliography{IEEEabrv,IEEEexample}

\end{document}